\documentclass[copyright]{eptcs}
\providecommand{\event}{J.~Dassow, G.~Pighizzini, B.~Truthe (Eds.): 11th International Workshop\\
on Descriptional Complexity of Formal Systems (DCFS 2009)} % Name of the event you are submitting to
\providecommand{\volume}{3}
\providecommand{\anno}{2009}
\providecommand{\firstpage}{147} %147--156
\providecommand{\eid}{14}
\usepackage{breakurl}        % Not needed if you use pdflatex only.

\input{dcfs.tex}

\usepackage{amsmath,amsfonts}
\usepackage{graphicx}
\usepackage{mathrsfs}

\newcommand{\inititem}{\setcounter{enumi}{0}\nextitem}
\newcommand{\nextitem}{\addtocounter{enumi}{1}(\theenumi)}

\newcommand{\up}{\mathchoice
  {\displaystyle\textsc{Up}}%
  {\textstyle\textsc{Up}}%
  {\textstyle\textsc{Up}}%
  {\textstyle\textsc{Up}}%
}
\newcommand{\down}{\textsc{Down}}
\newcommand{\lfam}{\mathscr{L}}
\newcommand{\resp}{respectively}

\begin{document}

\title{On Measuring Non-Recursive Trade-Offs}

\def\titlerunning{On Measuring Non-Recursive Trade-Offs}
\def\authorrunning{H.~Gruber, M.~Holzer, M.~Kutrib}

\author{Hermann Gruber \qquad Markus Holzer \qquad Martin Kutrib
\institute{Institut f\"ur Informatik --
  Universit\"at Giessen\\
  Arndtstra\ss e 2 -- 35392 Giessen -- Germany}
\email{$\{$gruber,holzer,kutrib$\}$@informatik.uni-giessen.de}
}

\maketitle

\begin{abstract}
  We investigate the phenomenon of non-recursive trade-offs between
  descriptional systems in an abstract fashion. We aim at categorizing
  non-recursive trade-offs by bounds on their growth rate, and show
  how to deduce such bounds in general.  We also identify criteria
  which, in the spirit of abstract language theory, allow us to deduce
  non-recursive tradeoffs from effective closure properties of
  language families on the one hand, and differences in the
  decidability status of basic decision problems on the other. We
  develop a qualitative classification of non-recursive trade-offs in
  order to obtain a better understanding of this very fundamental
  behaviour of descriptional systems.
\end{abstract}

\section{Introduction}
\label{sec:introduction}

In computer science in general, and also in the particular field of
descriptional complexity, we try to classify problems and mechanisms 
according to different aspects of their tractability. Often the first
distinction we make in such a classification is to check whether
a problem admits an effective solution at all. If so, we usually
take a closer look and analyze the inherent complexity of the problem.
But undecidable problems can also be compared to each other,
using the toolkit provided  by computability theory. Here, it turns
out that most naturally occurring problems are complete at some level  
of the arithmetic (or analytic) hierarchy. This has been a rather 
successful approach to understand the nature of many
undecidable problems we encounter in various computational settings.
As for decision problems, there are conversion problems between
different models that cannot be solved effectively. 
Indeed, they evade solvability {\em a forteriori} 
because the size blow-up caused by such a conversion cannot be 
bounded above by any recursive function. This phenomenon, 
nowadays known as \emph{non-recursive trade-off}, was 
first observed by Meyer and Fischer~\cite{Meyer:1971:edagfs}
between nondeterministic pushdown automata and finite automata.
Previously, it had been known that every deterministic 
pushdown automaton accepting a regular language 
can be converted into an equivalent finite automaton of at most
triply-exponential size. In contrast, Meyer and Fischer showed 
that if we replace ``deterministic pushdown automaton''
with ``nondeterministic pushdown automaton'',
then the maximum size blow-up can no longer be bounded 
by any recursive function.
Since that time there has been a steadily growing list of results where this
phenomenon has been observed,
e.\,g.,~\cite{%
Borchard:2002:nrtokfg,%
Goldstine:2002:dcmlr,%
Hartmanis:1980:sdrl:art,%
Hartmanis:1983:gsuslr,%
Herzog:1997:pdabnba,%
Kapoutsis:2004:kpotkdtonr:proc,%
kutrib:2005:dphcp:art,%
Malcher:2002:dccadq,%
Malcher:2004:dcia,%
Schmidt:1977:sducfl,%
Sunckel:2004:dcmcdgs:proc,%
Valiant:1976:nsddl%
}. 
In~\cite{kutrib:2005:pnrto:art} a survey is given that also presents
a few general proof techniques for proving such results. While it seems to
be clear that non-recursive trade-offs usually sprout at the wayside of
the crossroads of (un)decidability, in many cases proving such trade-offs 
apparently requires ingenuity and careful automata constructions.  
While apparently we cannot get rid of this altogether, 
here we identify general criteria
where non-recursive trade-offs can be directly read off, provided
certain basic (un)decidability results about the descriptional
systems under consideration are known. 
The present work aims at making the first steps in paralleling the successful 
development of the abstract theory of languages, 
and in building a theory with unified proofs 
of many non-recursive trade-off results appearing in the literature. 
Besides new proof techniques in this domain, the present work also 
aims to provide a finer classification of such non-recursive 
trade-offs, in a similar vein to what has been done in the 
classification of undecidable problems. 

The paper is organized as follows: in the next section we introduce
the necessary notation on descriptional systems and computability
theory. Then in Section~\ref{sec:upper-and-lower-bounds} we prove
bounds on the trade-off function~$f$ that serves as a least upper bound for
the increase in complexity when changing from a descriptor
in~$\mathcal{S}_1$ to an equivalent descriptor
in~$\mathcal{S}_2$. Here, it turns out that the complexity of the
problem of the $\mathcal{S}_2$-ness of $\mathcal{S}_1$ descriptors
influences the growth rate of~$f$. Finally, in
Section~\ref{sec:proof-scheme-non} we develop easy-to-apply proof
schemes that allow one to deduce non-recursive trade-offs by closure
properties of language families and differences in the decidability
status of basic decision problems.

\section{Preliminaries and definitions}
\label{sec:def}

We denote the power set of a set $S$ by $2^{S}$. The empty word is
denoted by~$\lambda$, the reversal of a word $w$ by~$w^R$, and for the
length of $w$ we write $|w|$. We use $\subseteq$ for \emph{inclusions}
and~$\subset$ for \emph{strict inclusions}.

We first establish some notation for descriptional complexity.  In
order to be general, we formalize the intuitive notion of a
representation or description of a family of languages. A
\emph{descriptional system} is a collection of encodings of items
where each item $D$ \emph{represents} or \emph{describes} a formal
language $L(D)$.  The encodings can be viewed as strings over some
alphabet.

\begin{definition} 
  A \emph{descriptional system} $\mathcal{S}$ is a recursive set of
  non-empty finite descriptors, such that each descriptor $D \in \mathcal{S}$
  describes a formal language $L(D)$, and if $L(D)$ is recursive (recursively
  enumerable), then there exists an effective
  procedure to convert~$D$ into a Turing machine that decides
  (semi-decides)~$L(D)$. 
\end{definition}

The \emph{family of languages represented (or described) by some
  descriptional system} $\mathcal{S}$ is 
$$\lfam(\mathcal{S}) =\{\,L(D)\mid D \in \mathcal{S}\,\}.$$
For every language $L$, the set of its
descriptors in the system $\mathcal{S}$ is $\mathcal{S}(L) =\{\,D \in
\mathcal{S} \mid L(D) =L\,\}$.

Now we turn to measure the \emph{size of descriptors}. 
From the viewpoint that
a descriptional system is a collection of 
encoding strings, the length of the strings is a natural measure of size. 
But in order to obtain a more general framework we consider a 
\emph{complexity (or size) measure} for~$\mathcal{S}$ to be a total, 
recursive mapping $c: \mathcal{S} \to \mathbb{N}$.
 
\begin{definition}
  Let $\mathcal{S}$ be a descriptional system. A \emph{complexity
    (size) measure} for~$\mathcal{S}$ is a total, recursive function
  $c: \mathcal{S} \to \mathbb{N}$ such that for any alphabet $A$, the
  set of descriptors in $\mathcal{S}$ describing languages over $A$ is
  recursively enumerable in order of increasing size, and does not
  contain infinitely many descriptors of the same size.
\end{definition}

We will call measures with these properties \emph{reasonable}.
Whenever we consider the relative succinctness of two descriptional
systems $\mathcal{S}_1$ and~$\mathcal{S}_2$, we assume the
intersection $\lfam(\mathcal{S}_1) \cap \lfam(\mathcal{S}_2)$ to be
non-empty.

\begin{sloppypar}
\begin{definition}
  Let $\mathcal{S}_1$ be a descriptional systems with complexity
  measure~$c_1$, and $\mathcal{S}_2$ be descriptional systems with
  complexity measure $c_2$. A total function \mbox{$f: \mathbb{N} \to
    \mathbb{N}$,} with $f(n) \geq n$, is said to be an \emph{upper
    bound} for the increase in complexity when changing from a
  descriptor in~$\mathcal{S}_1$ to an equivalent descriptor
  in~$\mathcal{S}_2$, if for all $D_1 \in \mathcal{S}_1$ with $L(D_1)
  \in \lfam(\mathcal{S}_2)$ there exists a $D_2 \in
  \mathcal{S}_2(L(D_1))$ such that 
  $$c_2(D_2) \leq f(c_1(D_1)).$$
\end{definition}
\end{sloppypar}

If there is no recursive upper bound, the trade-off is said to be 
\emph{non-recursive}. In other words, there are no recursive functions serving as
upper bounds. That is, whenever the trade-off from one descriptional 
system to another is non-recursive, one can choose an arbitrarily large
recursive function $f$ but the gain in economy of description 
eventually exceeds $f$ when changing from the former system to the latter.
So, a non-recursive trade-off exceeds any difference caused by applying two
reasonable complexity measures. 

In the sequel, if not otherwise stated, we always assume that there 
is a reasonable complexity measure~$c_i$ associated with any descriptional 
system $\mathcal{S}_i$.
We are interested in classifying non-recursive trade-offs
qualitatively. As it will turn out, the \emph{$\mathcal{S}_2$-ness of
$\mathcal{S}_1$ descriptors}, i.\,e., the problem
%\begin{itemize}
%\item 
given a descriptor $D_1\in\mathcal{S}_1$ does the language
  $L(D_1)$ belong to $\lfam(\mathcal{S}_2)$?,
%\end{itemize}
plays a central role in this task. We assume the reader to be
familiar with the basics of recursively enumerable sets and degrees as
contained in~\cite{Rogers:1967:trfec}. In particular we consider the
\emph{arithmetic hierarchy}, which is defined as follows:
\begin{eqnarray*}
  \Sigma_1 &=& \{\,L\mid\mbox{$L$ is recursively enumerable}\,\},\\
  \Sigma_{n+1} &=& \{\,L\mid\mbox{$L$ is recursively enumerable 
in some $A\in\Sigma_n$}\,\},
\end{eqnarray*}
for $n\geq 1$.  Here, a language~$L$ is said to be recursively
enumerable in some~$B$ if there is a Turing machine with oracle~$B$
that semi-decides~$L$. Let~$\Pi_n$ be the complement of~$\Sigma_n$,
i.\,e., $\Pi_n=\{\,L\mid\mbox{$\overline{L}$ is in
  $\Sigma_n$}\,\}$. Moreover, let $\Delta_n=\Sigma_n\cap \Pi_n$, for
$n\geq 1$. Observe that $\Delta_1=\Sigma_1\cap\Pi_1$ is the class of
all recursive sets. Completeness and hardness are always meant with
respect to many-one reducibilities~$\leq_m$, if not otherwise
stated. Let~$K$ denote the \emph{halting set}, i.\,e., the set of all
encodings of Turing machines that accept their own encoding.  For any
set~$A$ define $A'=K^A$ to be the \emph{jump} or \emph{completion}
of~$A$, where~$K^A$ is the \emph{$A$-relativized halting set}, which
is the set of all encodings of Turing machines with oracle~$A$
that accept their own encoding, and define $A^{(0)}=A$ and
$A^{(n+1)}=(A^{(n)})'$, for $n\geq 0$.  By Post's Theorem we have that
$\emptyset^{(n)}$ is $\Sigma_n$-complete ($\overline{\emptyset^{(n)}}$
is $\Pi_n$-complete, \resp) with respect to many-one reducibility, for
\mbox{$n\geq 1$}, where~$\emptyset^{(n)}$ is the \emph{$n$th jump}
of~$\emptyset$.  Moreover, note that \inititem~$A\in\Sigma_{n+1}$ if
and only if~$A$ is recursively enumerable in~$\emptyset^{(n)}$ and
\nextitem~$A\in\Delta_{n+1}$ if and only if~$A$ is recursive in, or
equivalently \emph{Turing reducible} to, the
jump~$\emptyset^{(n)}$. In this case we simply write
$A\leq_T\emptyset^{(n)}$, where~$\leq_T$ refers to Turing
reducibility. In the forthcoming we also use the above introduced
framework on Turing machines and reductions in order to compute
(partial) functions.

A more revealing characterization of the arithmetic hierarchy can be
given in terms of alternation of quantifiers. More precisely, a
language~$L$ is in~$\Sigma_n$, for $n\geq 1$, if and only if there
exists a \emph{decidable} $(n+1)$-ary predicate~$R$ such that
$$
L=\{\,w\mid \exists y_1\,\forall y_2\,\exists y_3\,\cdots\, Q\, y_n:
R(w,y_1,y_2,\ldots,y_n)\,\},
$$
where~$Q$ equals~$\exists$ if~$n$ is odd, and~$Q$ equals~$\forall$
if~$n$ is even. The characterization for languages in~$\Pi_n$, for
$n\geq 1$ is similar, by starting with a universal quantification and
ending with an~$\forall$ quantifier, if~$n$ is odd, and an~$\exists$
quantifier, if~$n$ is even.

\section{Bounds for non-recursive trade-offs}
\label{sec:upper-and-lower-bounds}

In this section we classify non-recursive trade-offs by given upper
and lower bounds. It will turn out, that whenever a non-recursive
trade-off between descriptional systems~$\mathcal{S}_1$
and~$\mathcal{S}_2$ exists, its (upper) bound is induced by the
property of verifying the~$\mathcal{S}_2$-ness of an~$\mathcal{S}_1$
descriptor, i.\,e., the problem of determining, whether for a given
descriptor $D\in\mathcal{S}_1$ the language $L(D)$ belongs to
$\lfam(\mathcal{S}_2)$. In order to make this more precise we need the
following theorem---observe, that by definition a descriptional
system is at most recursively enumerable:

\begin{theorem}\label{thm:S2ness-of-an-S1descriptor}
  Let~$\mathcal{S}_1$ and~$\mathcal{S}_2$ be two descriptional
  systems. The problem of determining for a given descriptor
  $D_1\in\mathcal{S}_1$ whether the language $L(D_1)$ belongs to
  $\lfam(\mathcal{S}_2)$, i.\,e., the $\mathcal{S}_2$-ness of
  $\mathcal{S}_1$ descriptors, can be solved in~$\Sigma_2$, if
  \emph{both}~$\mathcal{S}_1$ and~$\mathcal{S}_2$ are recursive. In
  case \emph{at least one} descriptional system is not recursive (but
  recursively enumerable) the problem can be solved in~$\Sigma_3$.
\end{theorem}

\begin{proof}
  The problem to determine whether for a given descriptor
  $D_1\in\mathcal{S}_1$ the language $L(D_1)$ belongs to
  $\lfam(\mathcal{S}_2)$ is equivalent to
  $$\exists D_2\in\mathcal{S}_2\,\forall w\in A^*: w\in L(D_1)\iff w\in
  L(D_2),$$ where~$A$ is the input alphabet of the devices under
  consideration. If both~$\mathcal{S}_1$ and~$\mathcal{S}_2$ are
  recursive, the logical formula $w\in L(D_1)\iff w\in L(D_2)$ is
  already
  a decidable $3$-ary predicate, since one can convert both
  descriptors~$D_1$ and~$D_2$ into Turing machines that decide the
  languages~$L(D_1)$ and $L(D_2)$, respectively. Hence, the problem can
  be solved in~$\Sigma_2$.
  
  If at least one descriptional system is not recursive (but recursively
  enumerable), we argue as follows: We rewrite the above
  characterization of the problem by
  $$\exists D_2\in\mathcal{S}_2\,\forall w\in A^*: [w\in L(D_1)
  \implies w\in L(D_2)]\wedge[w\in L(D_2)\implies w\in L(D_1)],$$ and
  replace the implications equivalently by
  $$\exists D_2\in\mathcal{S}_2\,\forall w\in A^*: [w\notin L(D_1)
  \vee w\in L(D_2)]\wedge[w\notin L(D_2)\vee w \in L(D_1)].$$ 
Then observe that $w\in L(D_1)$ ($w\notin L(D_1)$, \resp) can be
  verified if there is a time bound~$t$ (for every time bound~$t$,
  \resp) such that the word~$w$ is accepted (is not accepted, \resp) by~$M_1$ in
  at most~$t$ steps. Here~$M_1$ is the equivalent Turing machine
  effectively constructed from~$D_1$. A similar statement holds for
  $w\in L(D_2)$ and $w\notin L(D_2)$. Moving these quantifiers to the
  front by the Kuratowksi-Tarski algorithm~\cite{Rogers:1967:trfec}
  results in a $\Sigma_3$ characterization using a $4$-ary decidable
  predicate for the problem in question.
Thus, the problem can be solved in $\Sigma_3$.
\end{proof}

A closer look at the previous proof reveals that equivalence between
descriptors from~$\mathcal{S}_1$ and~$\mathcal{S}_2$ can be solved
in~$\Pi_1$ if \emph{both} descriptional systems are
recursive. Otherwise this equivalence problem belongs to~$\Pi_2$ (in
case \emph{at least one} descriptional system is not recursive). Thus, the upper bound on the equivalence
problem is one less in the level of unsolvability than the
$\mathcal{S}_2$-ness of $\mathcal{S}_1$ descriptors.

Next we deduce an upper bound on the trade-off between two
descriptional systems.

\begin{theorem}\label{thm:upper-bound}
  Let~$\mathcal{S}_1$ and~$\mathcal{S}_2$ be two descriptional
  systems. If both~$\mathcal{S}_1$ and~$\mathcal{S}_2$ are recursive,
  then there is a total function~$f:\mathbb{N}\to\mathbb{N}$ that
  serves as an upper bound for the increase in complexity when
  changing from a descriptor in~$\mathcal{S}_1$ to an equivalent
  descriptor in~$\mathcal{S}_2$, satisfying $f\leq_T\emptyset''$. In
  case \emph{at least one} descriptional system is not recursive (but
  recursively enumerable) the function~$f:\mathbb{N}\to\mathbb{N}$
  can be chosen to satisfy $f\leq_T\emptyset'''$.
\end{theorem}

\begin{proof}
  We only prove the statement for the case where both descriptional systems are
  recursive; the proof in case at least one descriptional system is
  not recursive (but recursively enumerable) follows along similar
  lines. In what follows we describe a Turing machine with
  oracle~$\emptyset''$ that computes
  a total function~$f$ that may serve as
  an upper bound for the increase in complexity when changing from a
  descriptor in~$\mathcal{S}_1$ to an equivalent descriptor
  in~$\mathcal{S}_2$. 

  Let $n\in\mathbb{N}$ be given. First determine the finite set
  $c_1^{-1}(n)$ of $\mathcal{S}_1$-descriptors, which can be
  effectively computed by the assumptions on~$c_1$, since 
  the set of descriptors in $\mathcal{S}_1$ is
  recursively enumerable in order of increasing size, and does not
  contain infinitely many descriptors of the same size.
  Then for each $D_1\in c_1^{-1}(n)$ we proceed as follows: If
  $L(D_1)$ is in $\lfam(\mathcal{S}_2)$, then we determine the value
  $$\min_{D_2\in\mathcal{S}_2}\{\,c_2(D_2)\mid L(D_2)=L(D_1)\,\}$$ and
  store it in a list.  By the previous theorem and the fact that
  $\emptyset^{(n)}$ is $\Sigma_n$-complete
  ($\overline{\emptyset^{(n)}}$ is $\Pi_n$-complete, \resp) the
  question whether $L(D_1)\in\lfam(\mathcal{S}_2)$ can be answered by
  an $\emptyset''$ oracle. In case the answer is yes, we recursively
  enumerate the descriptors in~$\mathcal{S}_2$ in increasing order
  until we find one descriptor that is equivalent to~$L(D_1)$. Here
  the equivalence between descriptors from~$\mathcal{S}_1$
  and~$\mathcal{S}_2$ is checked by a query to an $\emptyset'$ oracle,
  which is one less in jump as the one used to verify the condition
  $L(D_1)\in\lfam(\mathcal{S}_2)$---see the remark after the previous
  theorem on the equivalence problem. This enumeration procedure
  terminates since we already know that
  $L(D_1)\in\lfam(\mathcal{S}_2)$.

  Finally, we also store the input value~$n$ in the list, and
  compute the maximum of all list elements, which can effectively be done
  since the list has only finitely many entries. This value is assigned
  to $f(n)$. By construction, the function~$f$ is total and serves as
  an upper bound for the increase in complexity when changing from a
  descriptor in~$\mathcal{S}_1$ to an equivalent descriptor
  in~$\mathcal{S}_2$. Moreover, since the described algorithm always
  terminates, we have shown that the function~$f$ is recursive in
  $\emptyset''$---our Turing machine asks queries to an $\emptyset''$
  and $\emptyset'$ oracle, but since the set~$\emptyset'$ is strictly
  less in the levels of unsolvability one can simulate these queries by
  appropriate~$\emptyset''$ questions. This shows the stated claim.
\end{proof}

What about lower bounds on the trade-off function~$f$? In fact, we
show that there is a relation between the function~$f$ and the
equivalence problem between~$\mathcal{S}_1$ and~$\mathcal{S}_2$
descriptors, in the sense that, whenever the former problem becomes
easy, the latter is easy too.

\begin{theorem}\label{thm:lower-bound}
  Let~$\mathcal{S}_1$ and~$\mathcal{S}_2$ be two descriptional
  systems and 
  $f:\mathbb{N}\to\mathbb{N}$ a total function that
  serves as an upper bound for the increase in complexity when
  changing from a descriptor in~$\mathcal{S}_1$ to an equivalent
  descriptor in~$\mathcal{S}_2$. Then we have:
  {\rm \begin{enumerate}
  \item {\it If both descriptional systems are recursive and 
    $f\leq_T\emptyset'$, then the $\mathcal{S}_2$-ness of
    $\mathcal{S}_1$ descriptors is recursive in~$\emptyset'$.}
  \item {\it If at least one descriptional system is not recursive
    (but recursively enumerable) and 
    $f\leq_T\emptyset''$, then the $\mathcal{S}_2$-ness of
    $\mathcal{S}_1$ descriptors is recursive in~$\emptyset''$.}
  \end{enumerate}}
\end{theorem}

\begin{proof}
  We only prove the statement if both descriptional systems are
  recursive. The proof in case at least one descriptional system is
  not recursive (but recursively enumerable) follows along similar
  lines. We construct a Turing machine with oracle~$\emptyset'$ that
  decides the $\mathcal{S}_2$-ness of $\mathcal{S}_1$ descriptors.

  Let~$D_1$ from the descriptional system~$\mathcal{S}_1$ be given.
  Since the total function~$f$ is an upper bound for the increase in
  complexity when changing from a descriptor in~$\mathcal{S}_1$ to an
  equivalent descriptor in~$\mathcal{S}_2$ we first compute
  $m:=f(c_1(D_1))$. For this purpose queries to oracle~$\emptyset'$
  are needed. In fact the Turing machine that realizes the Turing
  reduction from function~$f$ to~$\emptyset'$ is used as a sub-routine
  here. Then we determine the finite set $\{\,c_2^{-1}(k) \mid k\leq m\,\}$ of
  $\mathcal{S}_2$-descriptors, which can be done on a Turing machine in
  a finite number of steps due to the assumptions on the size
  measure~$c_2$. Then for each of these descriptors we check by
  asking oracle~$\emptyset'$ whether they are equivalent
  to~$D_1$---note that equivalence for $\mathcal{S}_1$ and
  $\mathcal{S}_2$ descriptors can be verified in~$\Pi_1$ and hence by
  oracle questions to~$\emptyset'$. If at least one equivalent
  $\mathcal{S}_2$-descriptor is found the Turing machine halts and
  accepts; otherwise the machine halts and rejects. This shows that
  the $\mathcal{S}_2$-ness of $\mathcal{S}_1$ descriptors is recursive
  in~$\emptyset'$, since the constructed Turing machine always halts.
\end{proof}

Now we are ready to show that only \emph{two} types of
non-recursive trade-offs within the recursively enumerable languages
exist! First consider the context-free grammars and the right-linear
context-free grammars (or equivalently finite automata) as
descriptional systems. Thus, we want to consider the trade-off between
context-free languages and regular
languages. In~\cite{Meyer:1971:edagfs} it was shown that this
trade-off is non-recursive. By Theorem~\ref{thm:upper-bound}, one can
choose the upper bound function~$f$ such that $f\leq_T \emptyset''$. On
the other hand, if $f\leq_T\emptyset'$, then by
Theorem~\ref{thm:lower-bound} we deduce that checking regularity for
context-free grammars is recursive in~$\emptyset'$ and hence belongs
to~$\Delta_2$. This is a contradiction, because
in~\cite{Cudia:1970:dhupfg} this problem is classified to be
$\Sigma_2$-complete. So, we obtain a non-recursive trade-off
somewhere in between $\emptyset''$ and $\emptyset'$, that is,
$f\leq_T \emptyset''$ but $f\not\leq_T \emptyset'$.

In order to obtain higher growth rates on the
upper bound function~$f$, we have to go beyond context-free
languages. When considering the trade-off between the descriptional
system of Turing machines and finite automata we are led to the
following situation. Since one of the descriptional systems is not
recursive (but recursively enumerable) the function~$f$ can be be chosen
to satisfy $f\leq_T\emptyset'''$ by Theorem~\ref{thm:upper-bound},
but~$f$ cannot be simpler than $\emptyset''$ with respect to Turing
reducibility since otherwise
regularity for recursively enumerable languages would belong
to~$\Delta_3$, which contradicts the $\Sigma_3$-completeness of this
problem~\cite{Cudia:1970:dhupfg}. So, we obtain a non-recursive trade-off
somewhere in between $\emptyset'''$ and $\emptyset''$, that is,
$f\leq_T \emptyset'''$ but \mbox{$f\not\leq_T \emptyset''$.}

Our previous considerations can be summarized in a proof scheme for
non-recursive trade-offs. The statement reads as follows.

\begin{theorem}
  Let~$\mathcal{S}_1$ and~$\mathcal{S}_2$ be two descriptional
  systems. Then the 
  trade-off between~$\mathcal{S}_1$ and~$\mathcal{S}_2$ is
  non-recursive, if one of the following two cases applies:
  {\rm\begin{enumerate}
  \item {\it If both descriptional systems are recursive and the
    $\mathcal{S}_2$-ness of $\mathcal{S}_1$ descriptors is at least
    $\Sigma_2$-hard or}
  \item {\it at least one descriptional system is not recursive (but
    recursively enumerable) and the $\mathcal{S}_2$-ness of
    $\mathcal{S}_1$ descriptors is at least $\Sigma_3$-hard.}
  \end{enumerate}}
  \noindent Here hardness is meant with respect to many-one reducibility.
\end{theorem}

\begin{proof}
  We only prove the case when both descriptional systems are
  recursive. The other case follows by similar arguments.  Assume to
  the contrary that the trade-off between~$\mathcal{S}_1$
  and~$\mathcal{S}_2$ is recursive. Then there is a recursive, total
  function~$f$ which serves as an upper bound for the increase in
  complexity when changing from a descriptor in~$\mathcal{S}_1$ to an
  equivalent descriptor in the descriptional
  system~$\mathcal{S}_2$. Because~$f$ is a total recursive function we
  can mimic the proof of Theorem~\ref{thm:lower-bound} which shows
  that in our setting the $\mathcal{S}_2$-ness of $\mathcal{S}_1$
  descriptors is recursive in~$\emptyset'$. Thus, it belongs
  to~$\Delta_2$, which contradicts our prerequisites, which states
  that this problem is $\Sigma_2$-hard. Thus function~$f$ is
  non-recursive.
\end{proof}

Finally, it is worth mentioning that the presented approach to
measure non-recursive trade-offs nicely generalizes to higher degrees
of unsolvability than recursiveness and recursively enumerability
leading to non-recursive trade-offs of arbitrary growth rate. To this
end, the definition of descriptional systems has to be generalized in
order to cope with languages classes of the arithmetic hierarchy in
general. Then the proofs of Theorems~\ref{thm:upper-bound}
and~\ref{thm:lower-bound} obviously generalize to this setting as
well. The tedious details are left to the interested reader.

\section{Proof schemes for non-recursive trade-offs}
\label{sec:proof-scheme-non}

This section is devoted to the question of how to prove
non-recursive trade-offs. Roughly speaking, most of the proofs 
appearing in the literature are basically relying on one of two
different schemes---see, e.\,g.,~\cite{kutrib:2005:pnrto:art}.  One of
these techniques is due to Hartmanis~\cite{Hartmanis:1980:sdrl:art},
which he subsequently generalized in~\cite{Hartmanis:1983:gsuslr}. 
Next we present two rather abstract methods for proving 
non-recursive trade-offs. In contrast to previous schemes, 
here we only use properties that are known from the literature
for many descriptional systems: these concern the decidability 
of basic decision problems on the one hand, and closure properties 
familiar from the study of abstract families of languages on the 
other hand.

To this end, we define effective closure of descriptional systems under
language operations. We illustrate the definition by example of
language union: Let $\mathcal S$ be a descriptional system. We say
$\mathcal S$ is {\em effectively closed under union}, 
if there is an effective construction that, given
some pair of descriptors $D_1$ and $D_2$ from~$\mathcal S$,
yields a descriptor from $\mathcal S$ for $L(D_1)\cup L(D_2)$.
Effective closure under other language
operations is defined in a similar vein. The system $\mathcal S$ is
effectively closed under intersection with regular sets,
if there is an effective procedure that, given a 
descriptor $D$ from $\mathcal S$ and a regular
language~$R$, constructs a descriptor from~$\mathcal S$ describing the set 
$L(D)\cap R$.
A descriptional system is called an {\em effective trio}, 
if it is effectively closed under $\lambda$-free morphism, 
inverse morphism and intersection with regular languages. 
If it is also effectively closed under general morphism, we speak of
an {\em effective full trio}. Every trio is also effectively closed
under concatenation with regular sets.

The proofs that follow are based on \emph{Higman-Haines sets} of languages.
These are the closures of a language~$L$ under the scattered subword
and superword relations. More formally, let $\le$ denote the partial
order on words given by the scattered subword relation, i.\,e., $v\leq
w$ if and only if $v=v_1v_2\cdots v_k$ and \hbox{$w=w_1v_1w_2v_2 \cdots
w_kv_kw_{k+1}$}, for some integer~$k$, where~$v_i$ and~$w_j$ are
in~$A^*$, for $1\leq i\leq k$ and $1\leq j\leq k+1$.  Then for a
language $L\subseteq A^*$, the set $\down(L)$ is defined as $\{\,x
\mid \exists y \in L\,:\, y \le x\,\}$, and the set $\up(L)$ as $\{\,x
\mid \exists y \in L\,:\, x \le y\,\}$. What makes these sets
extremely useful are the two facts that the Higman-Haines sets of {\em
  any} given set of words are regular~\cite{Ha69,Hi52}, and that the
closure properties enjoyed by full trios imply closure under taking
Higman-Haines sets:

\begin{lemma}\label{lem:effective-trio-higman}
  Let $\mathcal S$ be an effective trio. Then $\mathcal S$ is
  effectively closed under the operation $\up$. Furthermore, if
  $\mathcal S$ is an effective full trio, then $\mathcal S$ is also
  effectively closed under the operation $\down$.
\end{lemma}

\begin{proof}
  It is well known that trios are closed under substitution with
  $\lambda$-free regular sets, and that full trios are closed under 
   substitution with regular sets, see,
  e.\,g.,~\cite{Hopcroft:1979:itatlc:book}.  Observe that the proof
  immediately leads to an effective construction. For
  any set $L\subseteq A^*$, we obtain $\up(L)$ \textit{via} the
  $\lambda$-free regular substitution given by $a \mapsto A^*aA^*$
  for each $a\in A$, and we obtain the set $\down(L)$ via the
  substitution given by $a \mapsto \{\lambda,a\}$, for each $a\in A$.
\end{proof}

The proof of the next theorem is based on the operation $\down$.

\begin{theorem}\label{thm:proof-scheme-down-set}
 Let~$\mathcal{S}_1$ and~$\mathcal{S}_2$ be two descriptional
 systems that are effective full trios.
If 
 {\rm\begin{enumerate}
 \item {\it the infiniteness problem for~$\mathcal{S}_1$ 
  is not semi-decidable and}
 \item {\it the infiniteness problem for $\mathcal{S}_2$ is decidable,}
 \end{enumerate}}
 \noindent then the trade-off between~$\mathcal{S}_1$ and~$\mathcal{S}_2$ is 
 non-recursive.
\end{theorem}

Before we prove this theorem observe that the full trio conditions imply
that $\lfam(\mathcal{S}_1)\cap \lfam(\mathcal{S}_2)\supseteq \mathrm{REG}$, 
see, e.\,g.,~\cite{Hopcroft:1979:itatlc:book} for a proof of this fact.

\bigskip

\begin{proof}
  Assume to the contrary that the trade-off 
  between~$\mathcal{S}_1$ and~$\mathcal{S}_2$ is
  bounded by some recursive function~$f$. Then we argue as follows:
  Let $D\in \mathcal{S}_1$. Since~$\mathcal{S}_1$ 
  is an effective full trio, by Lemma~\ref{lem:effective-trio-higman} 
  one can effectively construct
  a $D'\in \mathcal{S}_1$ satisfying $L(D')=\down(L(D))$. 
  Since~$L(D')$ is regular and $\mathcal{S}_2$ contains 
  all regular sets, our assumption implies that there is an 
  equivalent descriptor of size at most $f(c_1(D'))$. 
 
  With the help of the conditions imposed on~$\mathcal{S}_2$, we
  can determine the set $F$ of all descriptors in~$\mathcal{S}_2$ 
  of size at most $f(c_1(D'))$ that describe only finite languages. 
  Note in particular that this set $F$ of descriptors is finite.
  Furthermore, we can determine 
  the length~$k$ of the longest word contained in any 
  of the languages denoted by descriptors in $F$ as follows:
  By effective closure under concatenation with regular sets, 
  and under intersection with regular sets, 
  we simply search for the largest~$k$ such that the language 
  $a^*\cdot(L(D_i) \cap \{\,w\in A^*\mid |w|\geq k\,\})$,
  which is in $\lfam(\mathcal{S}_2)$, 
  is still infinite. Here $a$~is an arbitrary alphabet symbol.
  
  Now we make use of the observation from~\cite{GHK07a} that
  $L(D)$ is finite if and only if $L(D')\!=\!\down(L(D))$ is finite;
  and infiniteness of the latter can be proved by finding a word
  in $L(D')$ that is larger than $k$. We construct a Turing machine
  accepting $L(D')$ from $D'$, and we simulate the Turing 
  machine on all inputs of length at least $k$ by dove-tailing.
  If~$L(D)$ is infinite, eventually one of these simulations will
  accept, and this semi-decides infiniteness. But this contradicts our
  assumption, because by Condition~(I) the family of 
  descriptors~$\mathcal{S}_1$ has a non-semi-decidable 
  infiniteness problem.
\end{proof}

Notice that the above conditions in particular 
imply that the emptiness problem for $\mathcal{S}_2$ is 
decidable. A similar proof works if we drop the requirement 
on~$\mathcal{S}_2$ being a full trio and impose instead the 
following slightly weaker conditions, which are more bulky 
to state: first, that it describes all regular sets, second that 
it is effectively closed under intersection with regular sets, third 
it is effectively closed under concatenation with regular sets,
and fourth that 
emptiness is decidable for~$\mathcal{S}_2$.

Next we list some applications. 
Indexed grammars, which appear in the statement of the next theorem, 
were introduced in~\cite{Ah68}, and ET0L systems were studied in,
e.\,g.,~\cite{RoSa80}. 

\begin{theorem}
  The following trade-offs are non-recursive:
  {\rm\begin{enumerate}
  \item {\it Between Turing machines and finite automata,}
  \item {\it between Turing machines and (linear) context-free grammars,}
  \item {\it between Turing machines and ET0L systems, and}
  \item {\it between Turing machines and (linear) context-free indexed
    grammars.}
  \end{enumerate}}
\end{theorem}

\begin{proof}
  It is well known that the finite automata, the context-free
  grammars, and the Turing machines each form an effective full
  trio~\cite{Hopcroft:1979:itatlc:book}. Also the indexed grammars as
  well as ET0L systems form an (effective) full trio, as proved
  in~\cite{Ah68} and~\cite{RoSa80}, by means of effective
  constructions. That the infiniteness problem for Turing machines is
  not semi-decidable is folklore, while infiniteness for the other
  language families under consideration is decidable---see the
  aforementioned references.
\end{proof}

The proof of our next theorem is based on the operation $\up$.
Here we need not require that the effective trios are full, but now
both must have decidable word problems.

\begin{theorem}\label{thm:proof-scheme-up-set}
 Let~ $\mathcal{S}_1$ and~ $\mathcal{S}_2$ be two 
 descriptional systems that are effective trios.
If
 {\rm \begin{enumerate}
 \item {\it $\mathcal{S}_1$ %is an effective trio and 
  has a decidable word problem but an undecidable emptiness problem, and}
 \item {\it $\mathcal{S}_2$ %is an effective trio and 
  has a decidable emptiness problem,}
 \end{enumerate}}
 \noindent then the trade-off between~$\mathcal{S}_1$ 
 and~$\mathcal{S}_2$ is non-recursive. 
\end{theorem}

Observe, that the trio conditions imply that the intersection of
$\lfam(\mathcal{S}_1)$ and 
$\lfam(\mathcal{S}_2)$ contains all $\lambda$-free regular sets
(cf.~\cite{Hopcroft:1979:itatlc:book}).

\begin{proof}
%Remember that we always may assume that 
%$\lfam(\mathcal{S}_1)\cap\lfam(\mathcal{S}_2)$ is
%non-empty, and 
Assume to the contrary that the trade-off 
between~$\mathcal{S}_1$ and~$\mathcal{S}_2$ is
bounded by some recursive function~$f$. 
Then we argue as follows:
Let $D\in \mathcal{S}_1$. By Condition~(I) one can effectively construct
a $D'\in \mathcal{S}_1$ satisfying $L(D') = \up(L(D)) \cap A^+$. 
Since~$L(D')$ is regular and $\mathcal{S}_2$ contains all 
$\lambda$-free regular sets, our assumption implies that
$L(D')$ has a descriptor in $\mathcal{S}_2$ of size at 
most $f(c_1(D'))$.

With the help of the conditions imposed on $\mathcal{S}_2$,
we can determine the set~$N$ of all descriptors in $\mathcal{S}_2$ 
of size at most $f(c_1(D'))$ that describe only non-empty languages.
Since $N$ is finite, we can write $N$ as $\{N_1,N_2,\ldots,N_n\}$.
Then for each~$i$ with $1\leq i\leq n$ determine the 
lexicographically first non-empty word~$w_i$ accepted by~$N_i$. 
Since $\mathcal{S}_2$ has a decidable emptiness problem, 
and it is an effective trio, the
word problem for ~$\mathcal{S}_2$ 
is also decidable.
So, this task can be accomplished by enumerating all words 
in increasing order and deciding the word problem for each word
and each remaining descriptor.  

Now we make use of the observation from~\cite{GHK07a} 
that~$L(D)$ is empty if and only if $L(D')=\up(L(D))$ is empty;
and the latter can be tested as follows: $L(D')$ is non-empty if
and only if
at least one of the words $w_i$ is in $L(D')$. 
Finally, we simulate the original descriptor~$D'$ on all~$w_i$'s 
by a terminating Turing machine, for $1\leq i\leq n$. 
If at least one of these words is accepted, 
then~$L(D)$ is non-empty, otherwise~$L(D)$ is empty. 
Thus, emptiness is decidable for~$\mathcal{S}_1$, 
a contradiction.%
\end{proof}

Finally, we list a few applications.  Growing context-sensitive
grammars, which appear in the statement of the next theorem, were
studied, e.\,g., in~\cite{BuLo92,DaWa86}. Observe that
context-sensitive grammars form an effective trio, and the decidability
status of the emptiness problem of these language families can be found
in the previously mentioned references. We skip the straight-forward
proof of the next theorem.

\begin{theorem}
  The following trade-offs are non-recursive:
  {\rm\begin{enumerate}
  \item {\it between growing context-sensitive grammars and finite automata,}
  \item {\it between growing context-sensitive grammars and (linear)
    context-free grammars},
  \item {\it between growing context-sensitive grammars and ET0L systems,}
  \item {\it between growing context-sensitive grammars and indexed grammars,}
  \item {\it between context-sensitive grammars and finite automata,}
  \item {\it between context-sensitive grammars and ET0L systems,}
  \item {\it between context-sensitive grammars and (linear) context-free 
  grammars,}
  \item {\it between context-sensitive grammars and indexed grammars.}
 \hfill\qed
  \end{enumerate}}
\end{theorem}

\bibliographystyle{eptcs}
\bibliography{gruber}

\end{document}